\documentclass[fleqn,usenatbib]{mnras}

\usepackage{newtxtext,newtxmath}

\usepackage[T1]{fontenc}

\DeclareRobustCommand{\VAN}[3]{#2}
\let\VANthebibliography\thebibliography
\def\thebibliography{\DeclareRobustCommand{\VAN}[3]{##3}\VANthebibliography}

\usepackage{graphicx}	
\usepackage{amsmath}	
\usepackage{amssymb}	

\title[Do gas-poor galaxy clusters have different galaxy populations?]{Do gas-poor galaxy clusters have different galaxy populations? The positive covariance of hot and cold baryons}

\author[E. Puddu et al.]{
Emanuella Puddu$^{1}$\thanks{E-mail: emanuella.puddu@inaf.it}
and Stefano Andreon$^{2}$
\\
$^{1}$INAF - Osservatorio di Capodimonte, Salita Moiariello 16, 80131 Napoli, IT\\
$^{2}$INAF - Osservatorio Astronomico di Brera, Via Brera, Milano, IT
}

\date{Accepted XXX. Received YYY; in original form ZZZ}

\pubyear{2021}

\begin{document}
\label{firstpage}
\pagerange{\pageref{firstpage}--\pageref{lastpage}}
\maketitle

\begin{abstract}
Galaxy clusters show a variety of intra-cluster medium properties at a fixed mass, among which gas fractions, X-ray luminosity and X-ray surface
brightness. In this work we investigate whether the yet-undetermined cause producing clusters of X-ray low surface brightness also 
affects galaxy properties, namely richness, richness concentration, width
and location of the red sequence,  colour, luminosity, and dominance of the brightest cluster galaxy.
We use SDSS-DR12 photometry and our analysis factors out the mass dependency to derive trends at fixed cluster mass. Clusters of low surface brightness for their mass have cluster richness in spite of their group-like luminosity. Gas-poor, low X-ray surface brightness, X-ray faint clusters for their mass, display 25\% lower richness for their mass at $4.4\sigma$ level. Therefore, richness and quantities depending on gas, such as gas fraction, $M_{gas}$, and X-ray surface brightness,  are covariant at fixed halo mass. In particular, we do not confirm the hint of an anti-correlation of hot and cold baryons at fixed mass put forth in literature. 
All the remaining optical properties show no covariance at fixed mass, within the sensitivities allowed
by our data and sample size.  
We conclude that X-ray and optical properties are disjoint, the optical properties not showing 
signatures of those processes involving gas content, apart from the richness-mass scaling relation. 
The covariance between X-ray surface brightness and richness is useful for an effective X-ray follow-up of low surface brightness clusters because it 
allows us to pre-select clusters using optical data of survey quality and prevent expensive X-ray observations.
\end{abstract}




\section{Introduction}
\label{intro}

Galaxy clusters trace the highest density peaks in the large-scale structure of the Universe. Their clustering and abundance provide powerful cosmological probes that can put constraints on the estimate of the cosmological parameters when the cluster mass is known \citep[][and references therein]{Vik2009,Allen2011,XMMLSS2018}.

Cluster masses are not directly observed, and the less indirect ones, such as
weak-lensing masses, require complex and not always practicable techniques \citep{okabe2010,hoekstra2012, melchior2015,stern2019}. It is, therefore, useful
to consider some observables  easier to measure, or a mass proxy, in order to define a mass-observable scaling relation \citep[e.g.][]{Vik2009,AeH,Bellag19,Mantz2018}, such as the stellar mass, the number of cluster galaxies, the X-ray luminosity, the temperature and the Sunyaev-Zel’dovich strength. 

Scaling relations exhibit a scatter, which is included as a parameter in the cosmological analysis because severely affects it. A larger scatter in such relations produces larger errors in the cosmological parameters estimates \citep{LimaHu2005}. 
The scatter can be due to the measurement errors of the involved quantities, but there is also an intrinsic component that reflects the variegated properties of the examined clusters sample \citep[][]{Variegate}; such properties are related to the formation and evolution scenarios \citep{Hartley08,Fujita18} and to the underlying physical processes taking place into  clusters \citep{Farahi20}.
Understanding the nature of the intrinsic component could contribute having a more deep insight into these processes \citep[for example AGN feedback, star formation, mergers, radiative cooling and other non-gravitational effects; e.g.][]{Voit05,Nagai06,Rowley04}. 
\citet{Mulroy19} and \citet{ZuHone11} highlighted the role of the central gas entropy in increasing the residual of cluster observables around a mean scaling relation. 
\citet{Green19} have shown that the hot gas morphological properties can be introduced in the mass–observable relations to reduce the scatter. Furthermore, if covariance were known between mass observables, such as X-ray luminosity, and richness,  it could be exploited to improve extraction of the cosmological parameters \citep{Cunha2009}.

In addition to the mass scaling relations scatter, the selection function is a crucial quantity that has to be estimated accurately in any cosmological study. If neglected, selection effects may bias the recovered scaling relation \citep[e.g.][]{Sandage88ARAA,AeH,AB2012,Finoguenov2020,kafer2019},
because some objects are altogether missing or under-represented. In such cases, the scatter is also underestimated.

Because X-ray luminous and centrally peaked clusters are more likely to be detected, it is generally acknowledged  \citep[e.g.][]{Pacaud2007,AM2011,ATP2011MNRAS,Xu18} that X-ray selected samples are biased towards these types of objects. The missing of the low-surface-brightness galaxy groups or clusters can affect the cluster cosmology:
from CMB and Planck measurements many more clusters are expected than those actually observed \citep{Planck2011}. 

To avoid the bias of the X-ray selection, \citet{Amazing} built the X-ray Unbiased Cluster Survey (hereafter XUCS) sample, which was selected independently of the intra-cluster medium (ICM) content.  
This cluster sample shows \citep{Variegate}, at a given mass, a larger scatter in X-ray luminosities and in gas fraction than those inferred in previous works on X-ray selected surveys \citep{Bo2007,Pratt2009,Planck2011}.

In this work, we want to explore the optical properties of the XUCS sample, looking for possible signatures in the optical bands for those clusters with 
large (or small) gas fraction, X-ray luminosity at a given mass, and X-ray mean surface brightness. 

The paper is organized as follows: in Sect.~\ref{thesample} we describe the sample; in Sect.~\ref{optical}, we derive and correlate with the cluster X-ray surface brightness optical parameters, such as richness, concentration, colour distribution; in Sect.~\ref{discussion} we discuss the results, and in Sect.~\ref{conclusion} we draw the conclusions.

Throughout the paper, we assume a $\Lambda$CDM model with $\Omega_M=$ 0.3, $\Omega_{\Lambda}=$ 0.7, and H$_0=$ 70 km Mpc$^{-1}$s$^{-1}$.

\section{The sample}
\label{thesample}

\begin{figure}
   \centering
   \includegraphics[width=8cm]{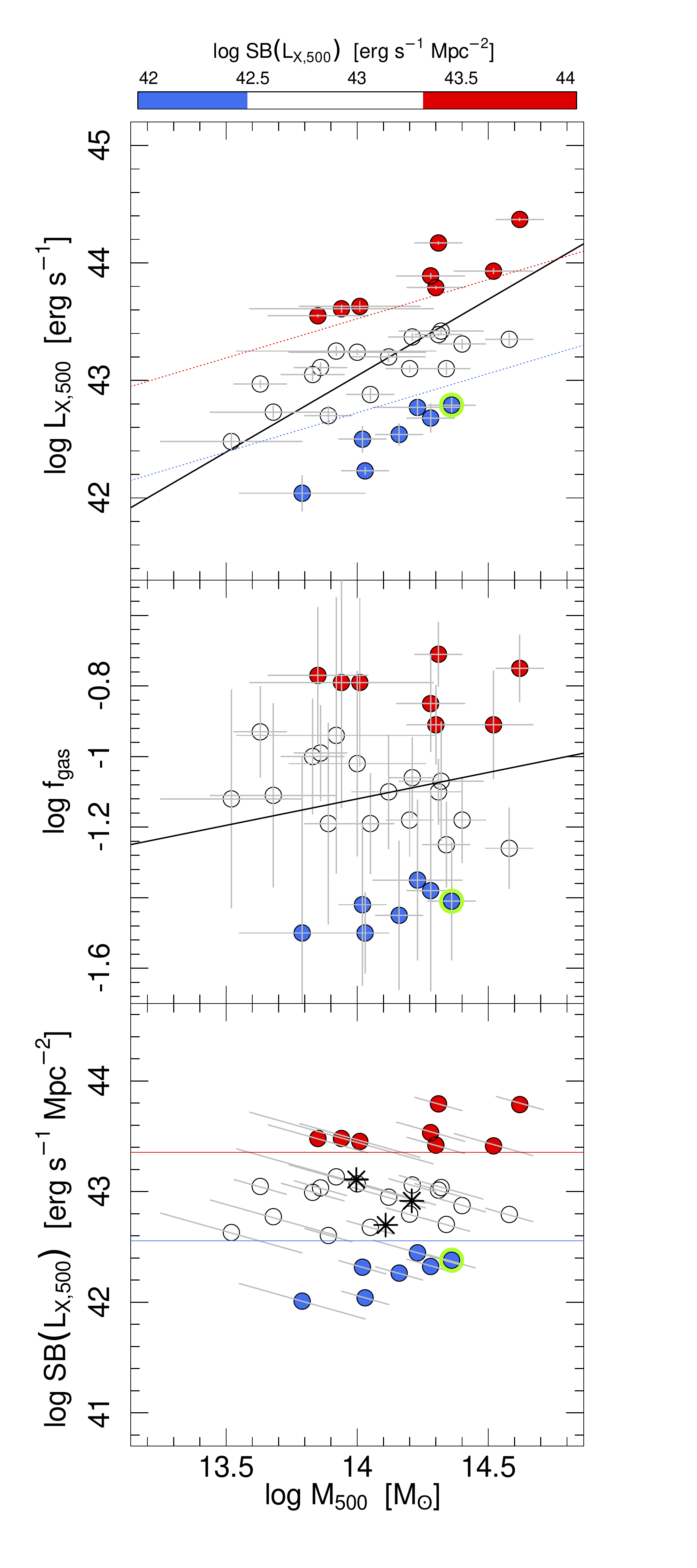}
   \caption{Gas mass and X-ray luminosity vs. mass for XUCS sample. 
   The two upper panels reproduce Fig.~4 of \citet{Variegate}: core-excised luminosity (upper panel) and gas fraction (central panel) vs. caustic mass. The solid lines indicate the mean fit respectively for the X-ray luminosity and the gas fraction slopes derived in \citet{Variegate}. In the lower panel, we plot X-ray surface brightness.
   Points are colour-coded by SB$_X$: we define as high-SB$_X$ clusters those above the red line and as low-SB$_X$ clusters those under the blue line. These red and blue lines are also displayed on the upper panel. 
   The most massive low-SB$_X$ cluster is CL2015, marked by a green circle. The clusters Abell~1631, Abell~76 and Abell~2399, which are low-SB$_X$ clusters in literature, overplotted to our sample (starred symbols).
   Clusters that are gas poor for their mass are also $L_X$-faint (for their mass).}
   \label{M500-DMgas}
\end{figure}

In this work, we use the XUCS sample of \citet{Amazing}.
This cluster sample  was extracted  
from the C4 catalogue of \citet{Miller05}, a catalogue built searching for overdensities in the seven-dimensional space of position, redshift and colours of the redshift survey performed by the Sloan Digital Sky Survey \citep[][]{Sloan}. 
XUCS clusters are selected among those C4 clusters with more than 50  spectroscopic members within 1 Mpc and consequently having reliable dynamical masses derived by velocity dispersion. Clusters with $\sigma_v>$ $500$ km s$^{-1}$ were selected, within an additional mass-dependent redshift range (0.050$< z <$0.135) introduced to facilitate {\it Swift} X-ray observations. 
The XUCS sample is composed of 34 clusters \citep{Variegate}, two of which were later excluded. The authors showed that this removal
does not alter the original properties of the sample.
The remaining 32 clusters, used in this work, are listed in Table \ref{table:1}. 
All the XUCS clusters have caustic masses \citep{Amazing}, derived using the caustic technique \citep{DG1997,diaferio99,Serra2011} which makes no assumption on the dynamical status.
The median number of members within the caustics is 116 and the inter-quartile range is 45. 
The mass range of the XUCS sample is 14 $<$log$M_{500}/M_{\odot} \leq $14.5. 
High-quality measurements of X-ray quantities were obtained in \citet{Amazing} from {\it Swift} observations or, for a few clusters, from available XMM-Newton or Chandra data: 
luminosity within $r_{500}$ ($L_{X,500}$) and core-excised ($L_{X,500,ce}$) luminosity; gas density profiles, gas mass  and gas fraction \citep[see][for details]{Amazing,Variegate}. 

In Fig.~\ref{M500-DMgas} we review some results found in the previous already cited works and useful for our analysis. 
In the upper panel, we reproduce
the Figure~4 of \citet{Variegate}, the $L_{X,500,ce} - M_{500}$ relation with the fit to the data (overplotted black line). 
In the middle panel, we show the 
$f_{gas}-M_{500}$ relation with the fit (black line) obtained by 
\citet{Variegate} (their Eq.~5). 
In the lower panel, we plot the mean
X-ray surface brightness derived by dividing the core-excise luminosity $L_{X,500,ce}$ by the area inside $r_{500}$, the latter derived from $M_{500}$.  

We define high/low surface brightness (hereafter SB$_X$)
clusters in the following way: $\log$\,SB(L$_{X,500}$)$\geq43.35$\,erg\,s$^{-1}$Mpc$^{-2}$ and $\log$\,SB(L$_{X,500}$)$\leq42.55$\,erg\,s$^{-1}$Mpc$^{-2}$ respectively, as shown in Fig.~\ref{M500-DMgas} (high-SB$_X$ clusters are above the red line and low-SB$_X$ clusters are under the blue line).
Points are colour-coded according to the X-ray surface brightness, allowing to disentangle high/low-SB$_X$  clusters and highlighting their behaviours with X-ray luminosity and gas mass. The blue dots represent low-SB$_X$ clusters, the red and empty dots depict respectively high and intermediate SB$_X$ clusters.
Clusters that are low-SB$_{X}$, are also gas poor and $L_X$-faint for their mass.  As apparent from Fig.1, we
might have divided clusters in the very same groups through a mass-dependent cut using either X-ray luminosity or gas fraction, whose errors have almost no covariance with mass.
The most massive low-SB$_X$ cluster is CL2015, which was studied in detail in \citet{Andreon2019}. 

A few clusters have been recently 
baptized as X-ray faint by the authors \citep{Bab18,Mitsu18}. None of them qualifies as low SB$_X$ according to our definition 
(Fig.~\ref{M500-DMgas} - lower panel), since the population unveiled by XUCS is of even lower surface brightness.

\section{Analysis}
\label{optical}

To investigate if the variety of ICM properties at a fixed mass can be related to differences in the optical properties of the XUCS clusters,
we checked whether the high/low-SB$_X$ status is correlated to the following optical properties:
richness, richness concentration, width
and location of the red sequence,  colour and luminosity of the brightest cluster galaxy and magnitude gap with the second brightest galaxy.

\subsection{Richness computation} 
\label{richnessderivation}
We compute richness closely following \cite{AeH}, \cite{AeC} and \cite{Making}. 
We extracted 3$^\circ\times$3$^\circ$ boxes around the cluster centre from the SDSS-DR12 photometric catalogue of galaxies \citep{DR12galcat}. 
In particular, we used the Sloan {\it cModel} magnitudes for the $g$ and $r$ bands, and computed the galaxy colour $g$-$r$ by means of the {\it Model} magnitudes. All the photometric quantities were corrected by extinction and the $g$-$r$ was also corrected for the slope of the cluster red-sequence computed at the $r$ magnitude corresponding to $M_{V,z=0}$=$-20$\,mag at the cluster redshift. As limiting magnitude, we considered $M_{V,z=0}$=$-20$\,mag, which is approximately the completeness magnitude of SDSS at $z=0.3$, whereas XUCS are much closer to us.

We counted red members within a specified luminosity range and colour, and within a given cluster-centric scale radius, in this case $r_{500}$.
The $g$-$r$ colour of the red sequence and the $r$ magnitude corresponding to $M_{V,z=0}$=$-20$\,mag were obtained by the \citet{BeC03} (hereafter BC03) model of passively evolving galaxies, considering a simple stellar population of solar metallicity with a Salpeter initial mass function (IMF) and a formation redshift of $3$. The expected colour, $(g-r)_{BC03}$, 
was corrected by an offset of $-0.08$ in order to account for the difference between the metallicity of $M_{V,z=0}$=$-20$\,mag galaxies and the assumed Solar metallicity of the model.
Then, we define red galaxies as those with $(g-r)_{BC03}-0.2<g-r< (g-r)_{BC03}+0.1$. 

\begin{figure}
   \centering
   \includegraphics[width=10cm]{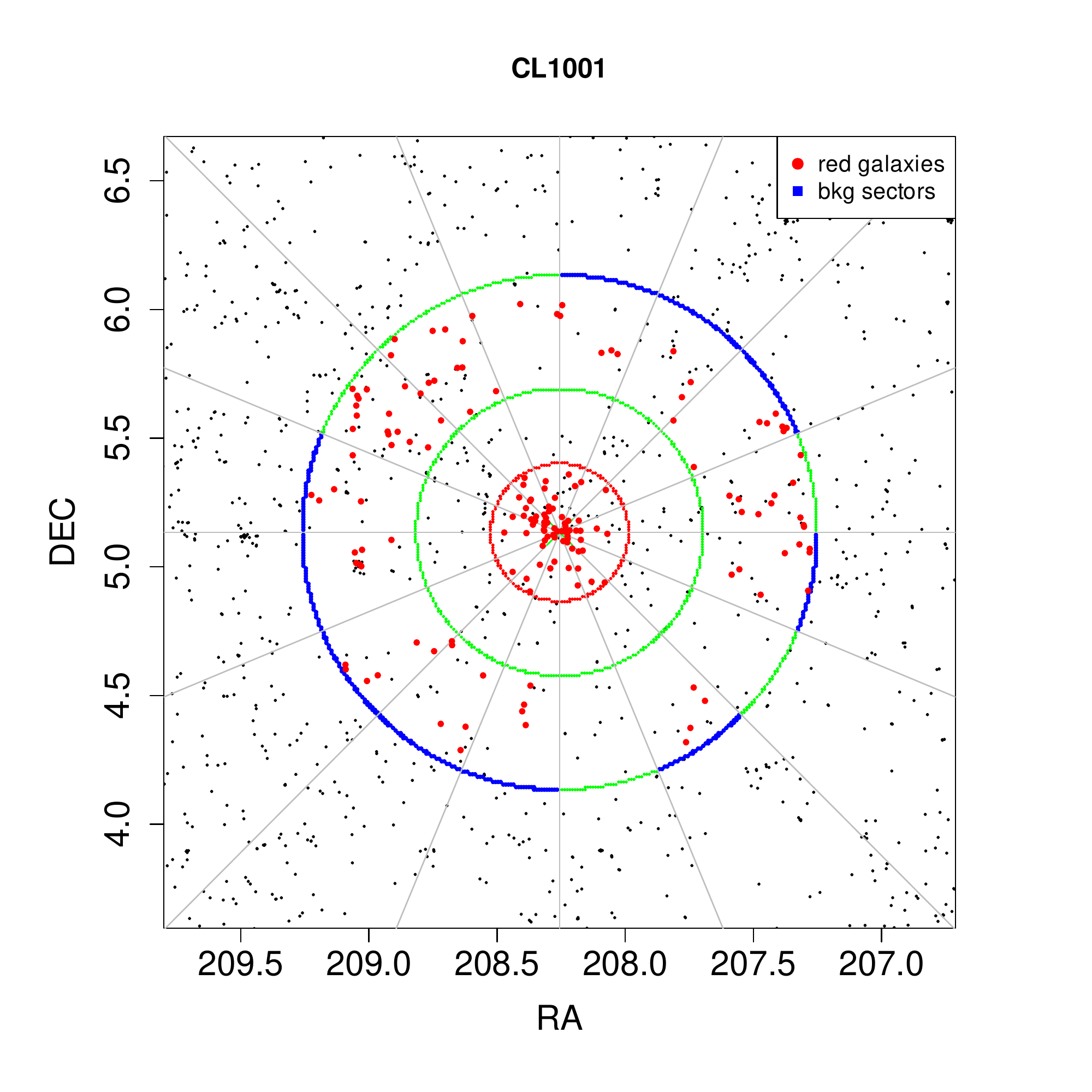}
   \caption{Diagnostic plot    
illustrating the richness determination for one example cluster (CL1001): the background contribution,
   estimated by counting galaxies falling within the 10 sectors (marked in blue) and the annulus (marked in green)
   is subtracted from the number of galaxies inside $r_{\Delta}$ (red circle around the centre), in this case $r_{200}$. }
   \label{richnesscomputation}
\end{figure}

In order to determine richness, the contribution from background/foreground galaxies has to be estimated and subtracted from the line of sight galaxy counts inside $r_{500}$. For the background estimate, we adopted an annulus with inner and outer radii of $3$~Mpc and $1$~degree. Then, we divided the annulus into 16 sectors and excluded from the background determination the three most and the three least populated sectors to allow for the presence of other
structures and problematic sectors, such as those occupied by bright stars or large galaxies. Fig.~\ref{richnesscomputation} illustrates the procedure for the cluster CL1001.

We computed $n_{500}$ as the difference between the galaxy counts inside $r_{500}$ and those in the selected 
background sectors normalized to the ratio of the relative solid angles. The richness within $r_{200}$ and half $r_{500}$, $n_{r_{500/2}}$, are similarly derived. 
Table~\ref{table:1} lists $n_{500}$ and $n_{r_{500/2}}$ values.

\begin{figure*}
   \centering
   \includegraphics[width=8cm]{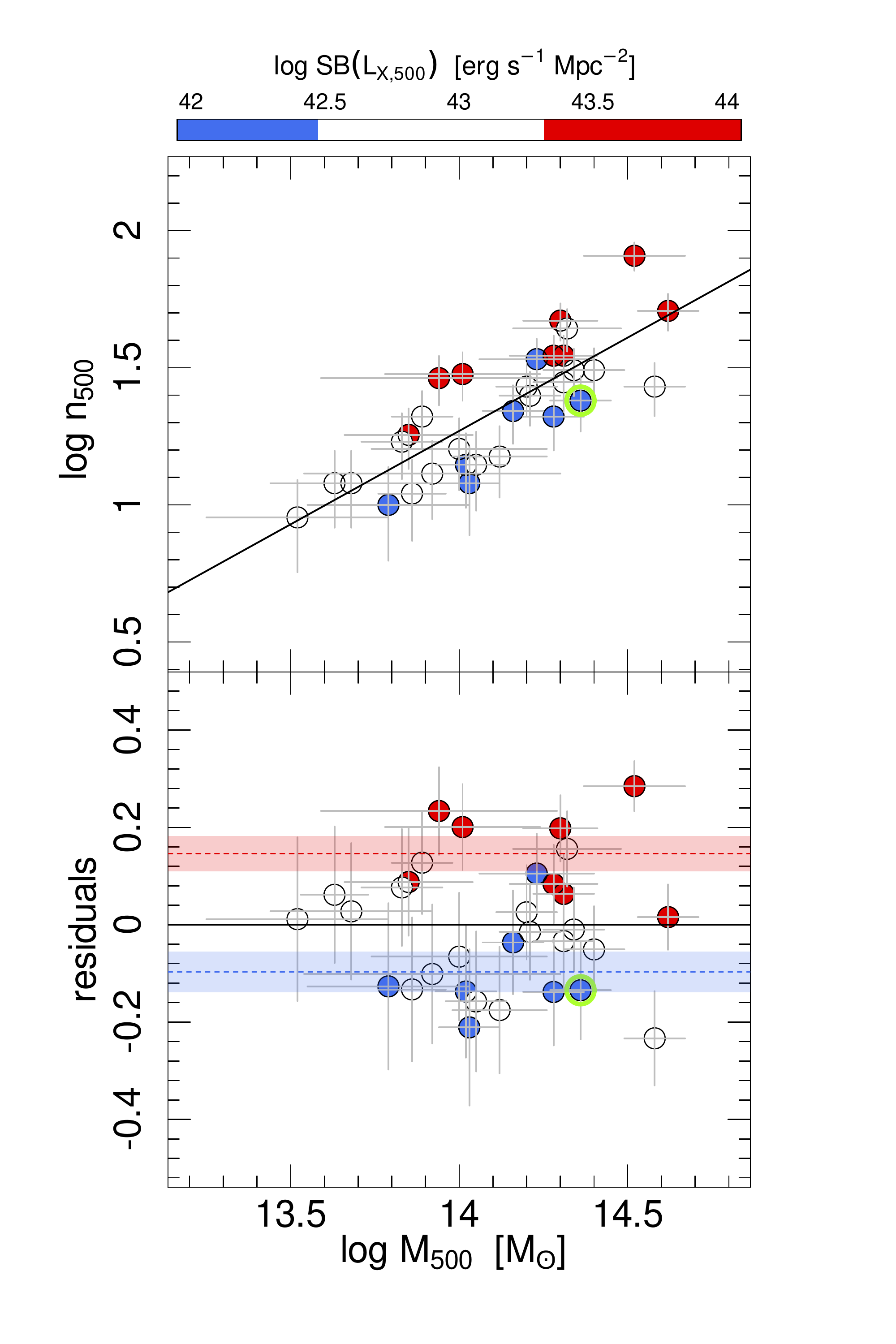}
   \includegraphics[width=8cm]{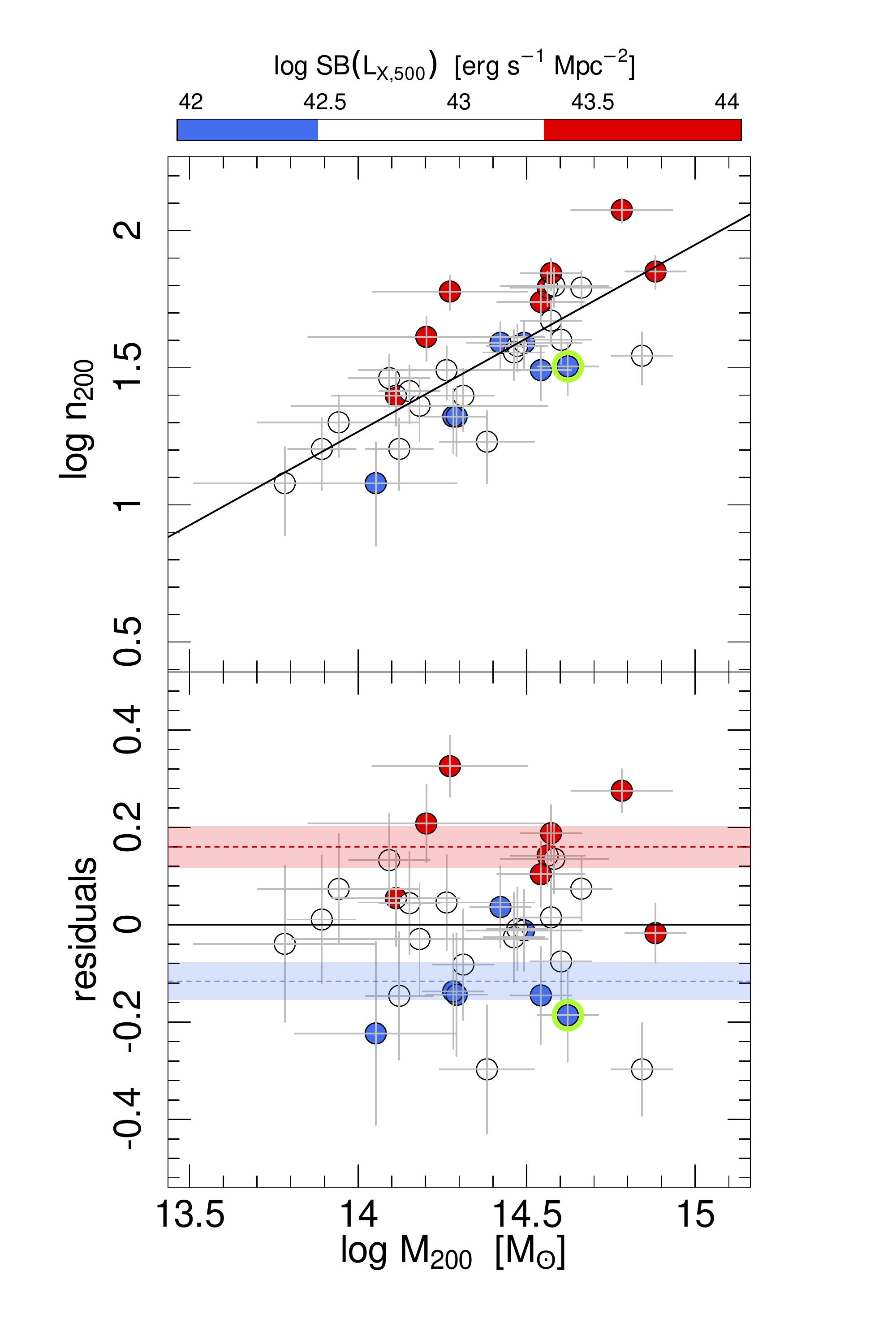}
 \caption{Richness, computed inside $r_{\Delta}$ (upper panels) vs. mass; the fit residuals are plotted in the lower panels. $\Delta$=$500$ at left and $\Delta$=$200$ at right.
 1$\sigma$ regions (shaded regions) around the mean values (dashed horizontal lines) are reported for the two populations. Points are colour-coded by SB$_X$ and CL2015 is marked by a green circle.} 
   \label{M500-richness}
\end{figure*}

\subsection{Differences in richness at fixed mass}
\label{trendonrich}

The left panel of Fig.~\ref{M500-richness} illustrates the dependence of richness $n_{500}$ on mass: in the upper panel, we plot the $n_{500}$-$M_{500}$ relation with its best fit (black line, ordinary least square fit), and in the lower panel, we plot the residuals. 
In addition to the well known correlation
telling that massive objects also have more galaxies,  high-SB$_{X}$ clusters (red dots) tend to lie above the best fit line (with offset$=$0.15$\pm$0.04) while low-SB$_{X}$ clusters (blue dots) are mostly below the best fit line (mean
offset$=$-0.10$\pm$0.04), as better visible in the bottom panel showing residuals. The two low/high-SB$_{X}$ populations are 4.4$\sigma$ apart. 
Clusters of low surface brightness for their mass have cluster richness despite  their group-like luminosity,
they have just about 25\% lower richness.

As already remarked, we might have identically divided the sample into three classes using an observable with errors showing no covariance with mass. Therefore, error covariance in the classification phase
plays no role in the observed covariance between richness and brightness at fixed mass.
Our $4.4\sigma$ significant result highlights a gas-optical richness covariance; at fixed mass: gas-poor clusters (which have low X-ray SB$_{X}$, see Fig.~1) are also poor for their mass. The right panel repeat the comparison using the ovedensity $\Delta$=$200$ in
place of $\Delta$=$500$, with similar results.

\subsection{Differences in richness concentration}
\label{rich_conc}

\begin{figure}
   \centering
   \includegraphics[width=8cm]{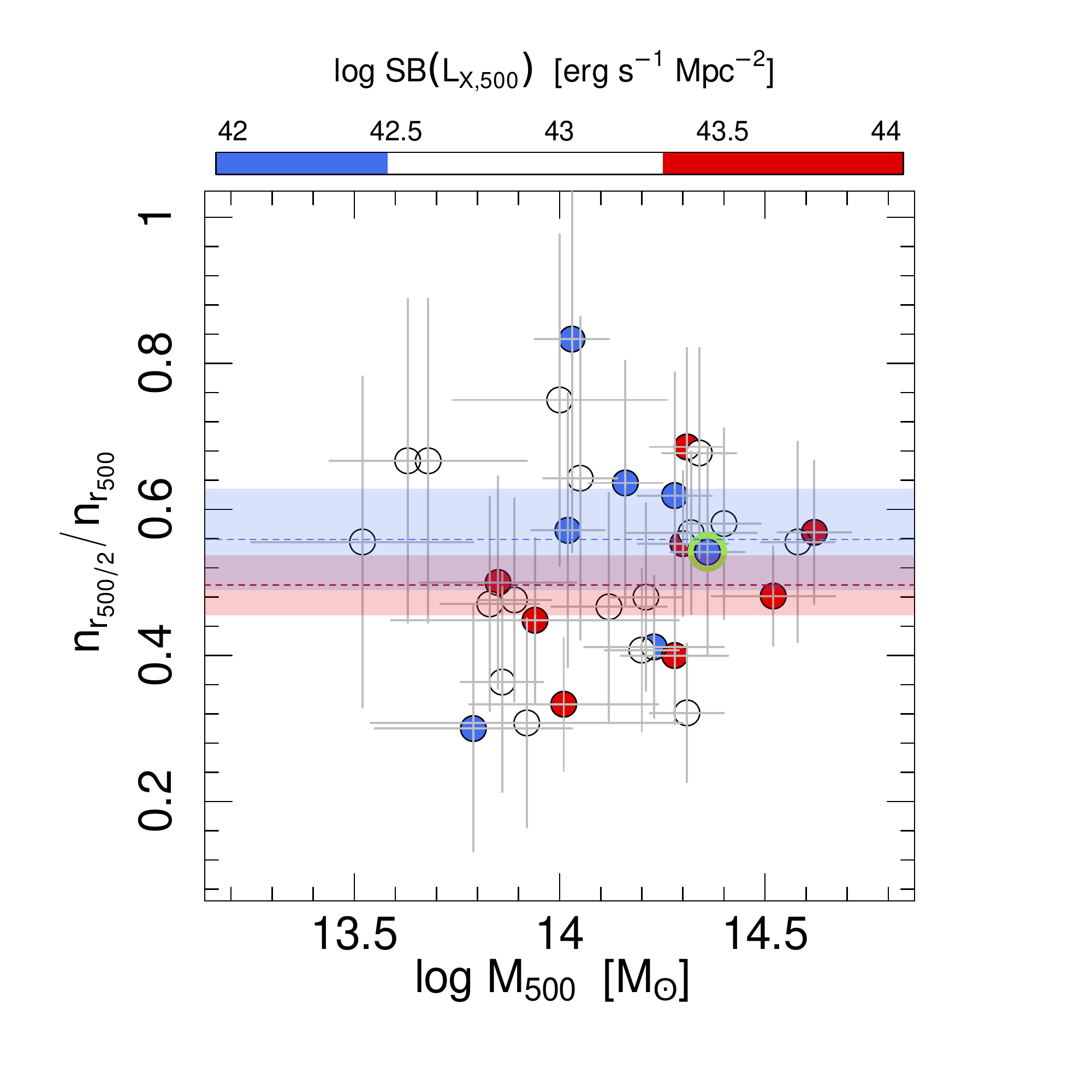}
   \caption{Galaxy concentration vs. mass. 1$\sigma$ regions (shaded regions) around the mean values (dashed horizontal lines) are reported for the two populations. Concentration increases upward.
   Clusters are colour-coded by SB$_X$ and CL2015 is marked by a green circle.
   }
   \label{M500-concentration}
\end{figure}

To measure the richness concentration, we computed the ratio between the number of galaxies inside half $r_{500}$ and the number of galaxies inside $r_{500}$, $n_{r500/2}/n_{r500}$,
shown in Fig.~\ref{M500-concentration}.
There is no obvious segregation between high-SB$_X$ and low-SB$_X$ clusters, since
the two mean concentrations ($0.50\pm0.04
$ and $0.56\pm0.07$, respectively) are $0.06$ away, corresponding to $\sim 0.78\sigma$.  
Given the sample size and the errors, we are only sensitive to a $0.22$ difference in concentration at $3\sigma$, comparable to the
spread of our data.

\subsection{Differences in width and location of the red sequence}
\label{sigmacol}

\begin{figure}
   \centering
   \includegraphics[width=8cm]{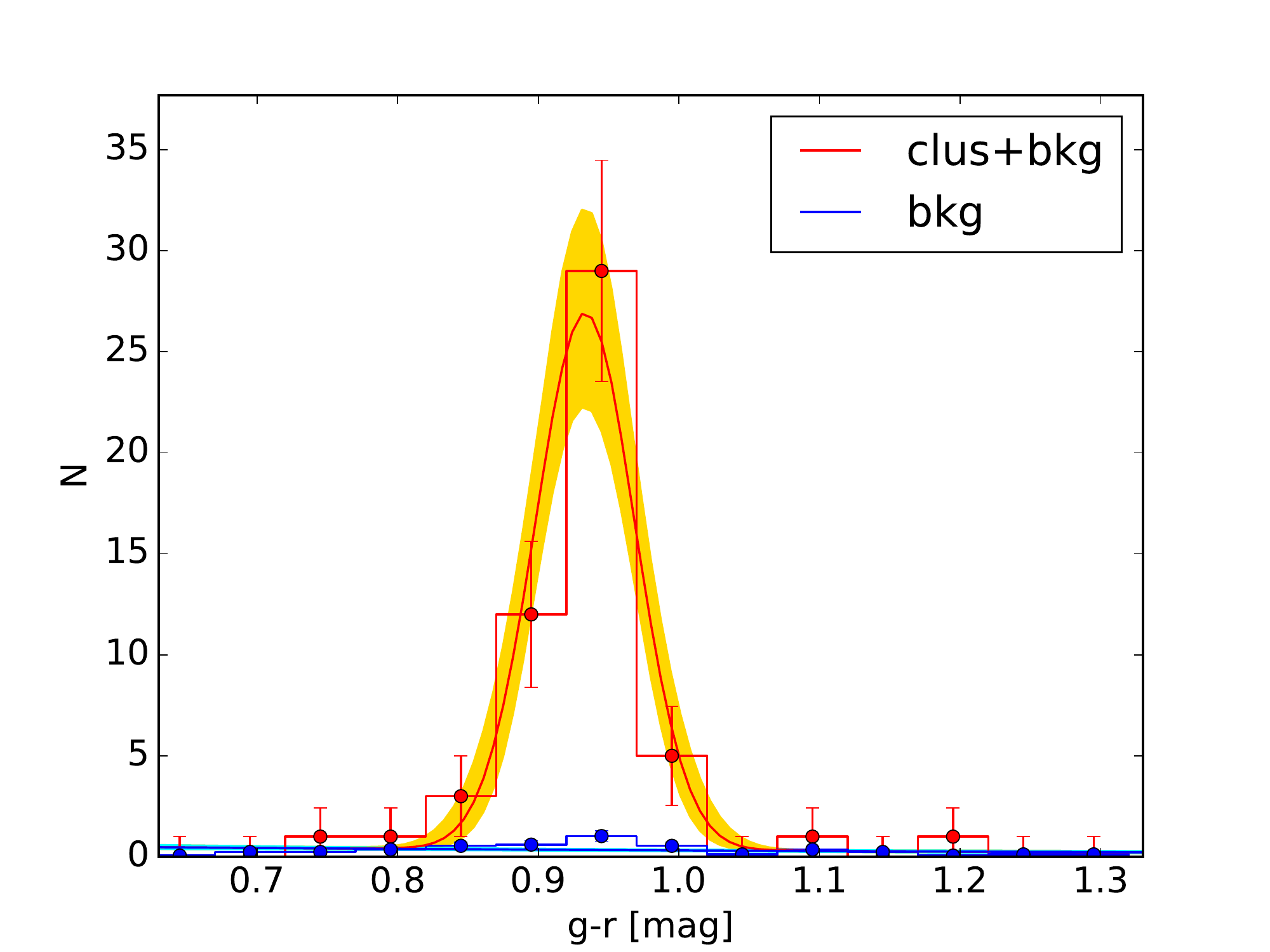}
   \caption{Illustration of red sequence width computing procedure for the cluster CL1001 as an example. 
The Gaussian+background fit (red curve), performed on unbinned counts, is overplotted to the binned red histogram/dots representing the cluster+background colour distribution; blue histogram/dots represent the binned background contribution superposed to the unbinned background fit (blue line). The yellow/cyan shadow indicates the 68\% error.}
   \label{fitGauss}
\end{figure}

Since galaxies in the cluster direction include both members and galaxies lying along the same line of sight, we need to account for fore/background galaxies in the computation of the width of the red sequence.
More precisely, we model the colour distribution of the cluster with a Gaussian and the background with a power-law of degree two; we assume a uniform prior for the model parameters, which we constrain by using the unbinned counts in the cluster and control field directions, following \citet{AP2008}. 
As an example, Fig.~\ref{fitGauss} shows the binned colour distribution of the cluster+background (red histogram/dots) and the background (blue histogram/dots) for the cluster CL1001, and the relative fitted functions on unbinned data (respectively red and blue curves). The derived
red sequence peak colour and dispersion ($\sigma_{g-r}$) are reported in Table~\ref{table:1}. 

\begin{figure}
   \centering
   \includegraphics[width=8cm]{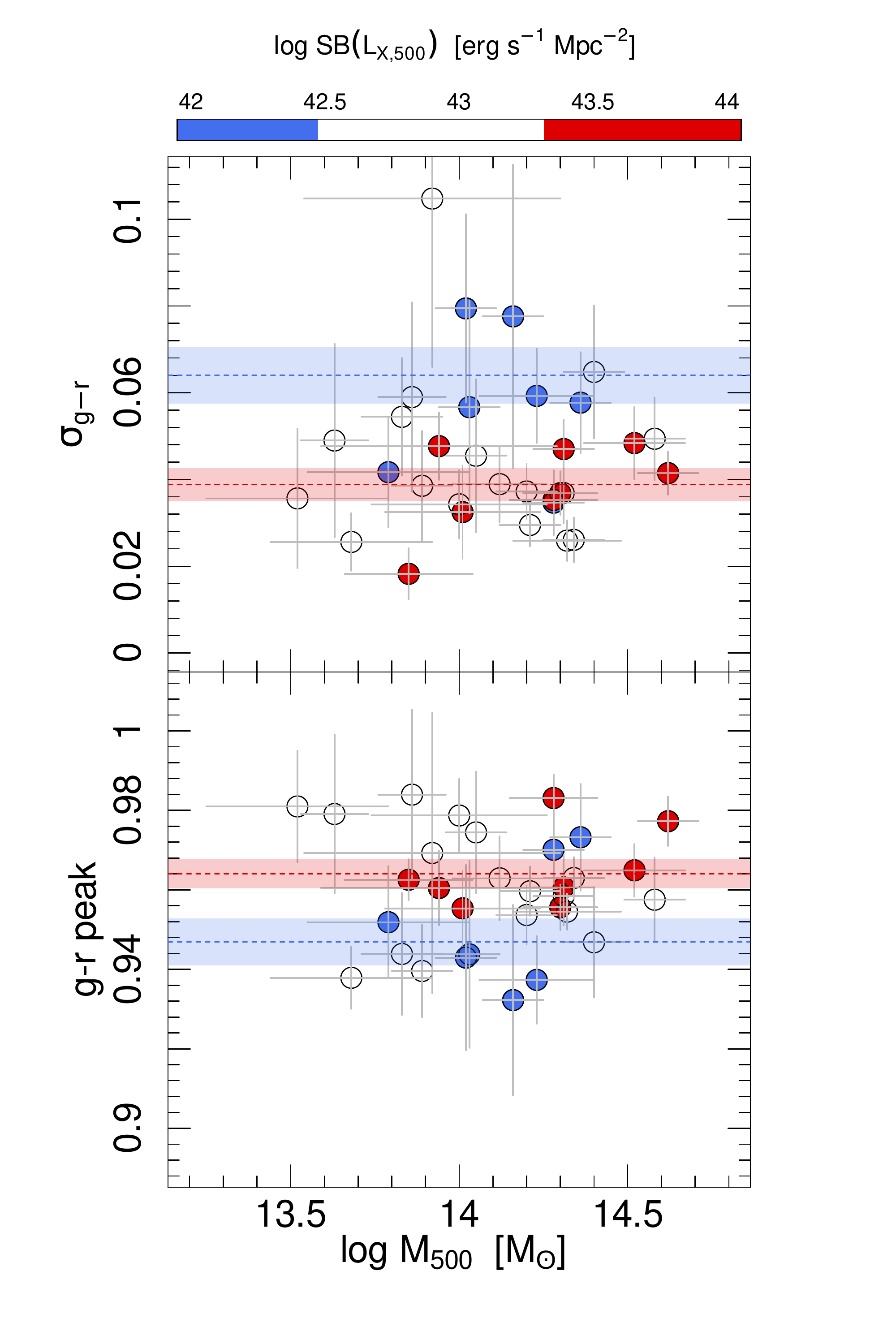}
   \caption{Width (upper panel) and location (lower panel) of the colour distribution vs. mass. Clusters are colour-coded by SB$_X$.
    The colour peak is redshift corrected at the survey mean. }
   \label{M500-sdcol}
\end{figure}

The upper panel of Fig.~\ref{M500-sdcol} shows the width $\sigma_{g-r}$ vs. $M_{500}$,  colour coded by SB$_X$.
The averages of the high-SB$_X$ and low-SB$_X$ populations ($0.039\pm0.004$ and $0.064\pm0.007$\,mag, respectively) are $0.025$\,mag away, about $3.29\sigma$
different. 

Since we have not modelled the colour distribution of cluster blue members, the
found width is inflated and made noisy by the possible presence of very few blue cluster members, which seems to be present for the two top blue points in Fig.~\ref{M500-sdcol}. Therefore, our widths should be read as upper limits of the red sequence width. We consider this result as tentative and a prompt for a future improved analysis where the blue cluster population is modelled.

The bottom panel of Fig.~\ref{M500-sdcol} shows the peak location ($g-r$ colour)
of the red sequence vs. $M_{500}$. We account for (minor) differences in redshift by correcting colours
at the mean sample redshift, $z=0.0879$ by using the BC03 model.
The average red sequence colour for the high-SB$_X$ and low-SB$_X$ populations are $0.964\pm0.004$ and $0.947\pm0.006$ mag respectively, about $0.017$\,mag away, or about $2.5\sigma$.
Therefore, there is no clear evidence of a relation between SB$_{X}$ and the red sequence location.
With our sample size and  errors, we are sensitive to $0.021$ mag differences in colour at $3\sigma$.

\subsection{Differences in BCG properties}
\label{BCGproperties}

\begin{figure*}
   \centering
   \includegraphics[width=8cm]{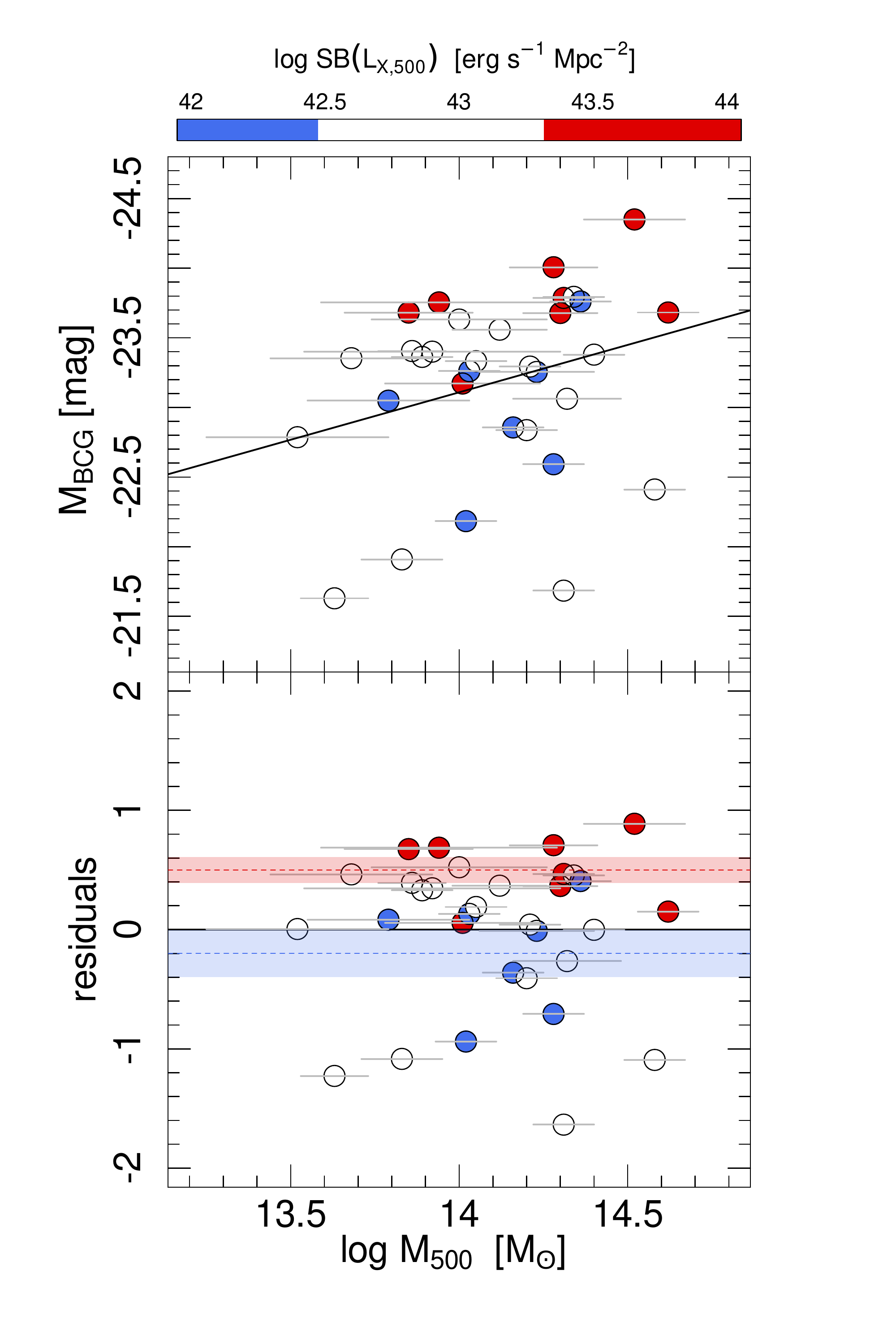}
   \includegraphics[width=8cm]{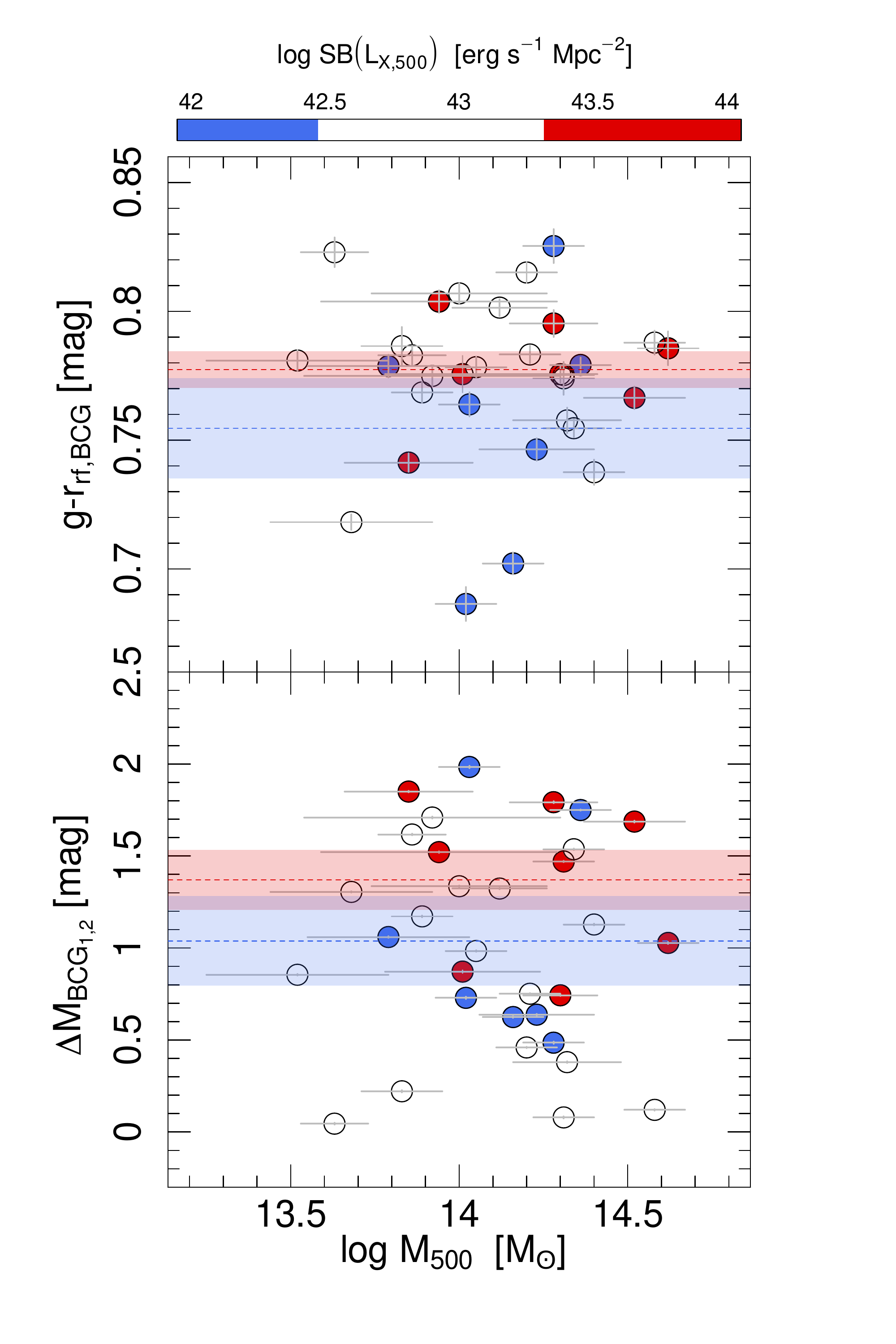}
   \caption{Left panels: absolute magnitude of BCGs vs. $M_{500}$ (top) with the best fit (black line) and the residuals (bottom). 
   Right panels: BCG rest frame colours (top) and magnitude difference between the BCG and the second brightest cluster galaxy ($\Delta M_{BCG_{1,2}}$, bottom) vs. $M_{500}$. Points are colour-coded by SB$_X$. The magnitude errors are smaller than the symbols. }
   \label{M500-MBCG}
 \end{figure*}

In this Section, we investigate if clusters with different SB$_X$ have BCGs 
with different properties, namely absolute magnitude, their rest frame colour, and the magnitude gap between the first and second brightest galaxy. 

The BCG was already identified in \citet{Amazing} as the brightest cluster galaxy on the red sequence
closest to the X-ray centre.
The second BCG is defined as the second brightest galaxy on the red sequence inside $r_{500}$. 
The top-left  panel of Fig.~\ref{M500-MBCG} shows the BCGs $r$-band absolute magnitude $M_{BCG}$ vs. $M_{500}$. The BCG absolute magnitude and cluster mass are slightly correlated ($\rho \sim 0.3$), although with a wide scatter. 
Most of the high-SB$_X$ clusters (red dots) tend to lay above the best fit line (black line, ordinary least square fit) with an average offset of $-0.50\pm0.11$, as illustrated in the bottom-left panel.  

At fixed cluster mass,
the luminosity of the dominant galaxy seems to be brighter in the high-SB$_X$ clusters than in the low-SB$_X$ ones, but only at about $3\sigma$. 

The right upper panel of Fig.~\ref{M500-MBCG} shows the BCG rest frame colour $g$-$r_{BCG}$ vs. $M_{500}$. 
The low-SB$_X$ mean value at $0.75\pm0.02$ is bluer than the high-SB$_X$ mean at $0.78\pm0.01$ of $1.14\sigma$; given our sample size and the errors, we are only sensitive to a $0.06$~mag difference in magnitude at $3\sigma$.

Finally, the bottom-right panel shows $\Delta M_{BCG_{1,2}}$, the difference in absolute magnitude between the first and second BCG, plotted vs. $M_{500}$. 
The mean values of $\Delta M_{BCG_{1,2}}$ for the low-SB$_X$ and high-SB$_X$ clusters are $1.04\pm0.24$ and $1.37\pm0.16$ respectively.
There are no significant differences in the magnitude gap between low-SB$_X$ and high-SB$_X$ clusters, our $3\sigma$ sensitivities being $0.9$~mag and the low/high SB$_X$ means being separated of $1.13 \sigma$. Our magnitude gap of CL2015 matches the value reported in \citet{Andreon2019}.

We can conclude that there are no strong evidences of relationship between the X-ray surface brightness and the parameters BCG colour or magnitude gap.


\section{Discussion}
\label{discussion}

It is a long established result that the most massive clusters have more of everything, that is, when the mass increases, the number of galaxies and the amount of gas mass also increase. This should not
be confused with our analysis at fixed mass.
In this work we 
find a covariance between richness and X-ray surface brightness at fixed total mass: at
a given mass the richer clusters are also those with more gas mass (more X-ray luminous or with higher SB$_{X}$), whereas the poorer clusters have also less gas mass. 
Fixing the mass, we can compare objects of the same class, with a similar formation history and formation time, and we can better disentangle the involved physical processes in a scenario less confused by different competing effects which dilute and blur each other. For example, \citet{2021arXiv210500068O} find indirect evidence of an SZ-faint cluster population, but the lack of mass measurements prevent them to address the nature of the faint population, whether they are intrinsically faint clusters or just line of sight projections, for example.

Relying on simulations, richness, formation time and magnitude gap are related to each
other.
The number of satellite galaxies in a halo, which is closely related to the cluster richness, depending on the formation time \citep{Farahi20}, because satellites in early formed haloes had more time to merge and reduce their number. Moreover, the cannibalism of the central galaxy, grown by major mergers, produces the lack of bright satellite galaxies \citep{Jones03,Giodini13} and consequently increases the magnitude gap between the first and the second BCG, which should therefore be indicative of the formation time.
Finally, earlier formed haloes should have deeper potential well and higher capability to retain the gas than later formed haloes \citep{Wechsler02,FarahiNat19}.
Following this scenario, we should expect at fixed halo mass a trend between richness, gas content and formation time.
However, in the data, we do find the expected covariance between richness and gas mass at
fixed halo mass, but
we do not find evidences for a relation between magnitude gap $\Delta M_{BCG_{1,2}}$ and SB$_{X}$ (Fig.~\ref{M500-MBCG}, right lower panel), indicating that our data does not support the hierarchical scenario described above or that competing effects may concur to blur the trend. 

Since less concentrated clusters have lower central densities, we also expect that less concentrated clusters are X-ray fainter, as CL2015 \citep{Andreon2019}. Our richness concentration measurements do not support this expectation; however, our large errors on richness concentration makes our results inconclusive. In particular, the low  mass concentration CL2015 cluster also studied here, has not low richness concentration (Fig.~\ref{M500-concentration}), possibly because of our large errors on richness concentration.
We plan to improve our concentration measurement in a forthcoming paper considering a wider galaxy population inclusive of less massive galaxies and to investigate the subject using
the Magneticum simulation \citep{Dolag16}.
An other promising way to give insight to the outlined scenario is that of \citet{monterodorta2021influence}, who point out the relevant role of the mass accretion history in delineating the cluster properties, using the Illustris-TNG300 magnetohydrodynamical simulations.

\subsection{The practical utility of the observed SB$_{X}$-richness covariance}

We found that at fixed mass, low-SB$_{X}$ clusters contain fewer galaxies than clusters of normal or high brightness. If one is interested in 
X-ray faint clusters for their mass, this correlation is
potentially useful to preselect 
low-SB$_{X}$ clusters, hence saving X-ray exposure time. Indeed,  
Fig.~\ref{M500-richness} shows that selecting clusters with negative residuals
in the mass-richness scaling, either at the overdensity $\Delta=500$  or $200$, is effective in reducing the sample, by a factor of 2 to 3 depending
on which precise threshold in X-ray surface brightness is chosen.
Richness measurements are inexpensive and the observational material to derive them is already available as SDSS \citep{redmapper2016}, Panstarrs \citep{chambers2019panstarrs1} and Legacy at NOAO  \citep{legacy19}, which already observed almost the whole sky at the appropriate depth, unlike currently available X-ray surveys that have insufficient depth to discriminate low-SB$_{X}$ clusters from the others.
For example, CL2015, the most massive X-ray faint cluster in our sample \citep{Andreon2019}, might have been chosen this way (point with the green contour in Fig.~\ref{M500-richness})
in place than from pointed X-ray observations of a larger sample.  The BCG luminosity and red-sequence width are also useful to filter out clusters of high surface brightness, but it is less effective in discriminating clusters of average surface brightness from those of low surface brightness (Fig.~\ref{M500-MBCG} and Fig.~\ref{M500-sdcol}.)

\subsection{Richness including blue galaxies}

Our richness computation excludes blue cluster members. To test the impact of this population, we recomputed richnesses for galaxies of all colours
more massive than the mass limit used thus far, the one corresponding to a luminosity $M_{V,z=0}$=$-20$\,mag for our adopted SSP. Mass is estimated from the magnitude and colour following \citet{Bell2003}. These new richnesses turns out to be very similar to those already computed: the ratio between the two  has a mean value of 1.04$\pm$0.02, 1.00$\pm$0.02 and 1.07$\pm$0.02 for the high, low and intermediate SB$_X$ clusters, respectively. 

\begin{figure}
   \centering
   \includegraphics[width=8cm]{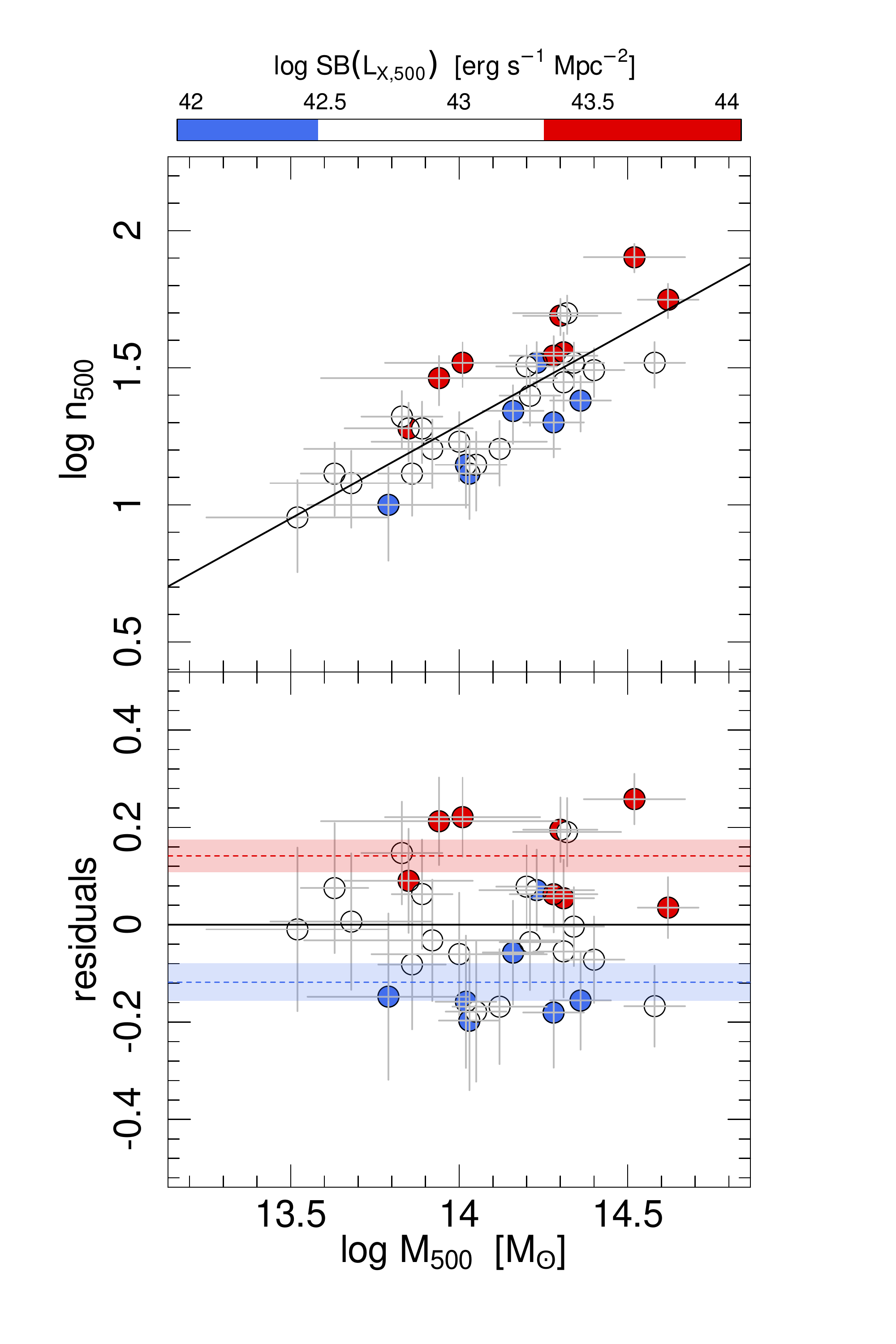}
   \caption{The same plot as in Fig.~\ref{M500-richness} but with the richness including the blue galaxies.}
   \label{M500-n500dimm}
 \end{figure}

Fig.~\ref{M500-n500dimm} shows the dependence of the newly defined richness on mass, and it is hardly distinguishable from the similar plot only
considering red galaxies in Fig.~\ref{M500-richness}.
Accounting for the blue population, high-SB$_X$ clusters (red dots) lie above the best fit line (with offset$=$0.14$\pm$0.03), and low-SB$_X$ clusters (blue dots) are mostly below the best fit line (mean offset$=$-0.12$\pm$0.04), as better visible in the bottom panel showing residuals. The two low/high-SB$_X$ populations are 5.1$\sigma$ apart.

\subsection{Covariance or anti-covariance?}

The sign of the covariance at fixed mass between mass proxies, such as richness or X-ray luminosity, plays a key role in the optimal extraction of cosmological parameters from cluster counts, especially in the case of mass proxies with
comparable scatter (Stanek et al. 2010). In such a case, positive covariance gives no improvement compared to the use of just one single proxy, uncorrelated proxies offer the expected $\sqrt{2}$ improvement, while anti-covariance may offer a five-fold improvement over the quality of each mass proxy.

We found a positive covariance between cold baryon (richness) and hot baryon (X-ray luminosity, surface brightness, gas fraction or gas mass) at fixed mass, in the sense that galaxies and gas go in the same direction: an excess of the former correspond to an excess of the latter, and we found it for a sample selected independently
of X-ray luminosity. By considering instead an X-ray selected sample, in which therefore over-bright clusters are overly-represented and clusters of low luminosity for
their mass are under-represented when not missing altogether, \citet{FarahiNat19}
found anti-covariance between hot and cold baryon \citep[namely X-ray luminosity and the near-infrared luminosity and richness, where the latter two quantities are tightly correlated,][]{Stellarmass} at fixed mass at $\sim2\sigma$ significance, in agreement
with the expectation that massive clusters are closed boxes. There are, however, observational evidences that clusters are not closed boxes: the total baryon fraction is not universal \citep[e.g. ][]{Stellarmass,Eckert2013} and 
the amount of metals in the intra-cluster medium is not the one expected given the observed galaxies \citep[e.g. ][]{RenziniAndreon14,ghizzardi2010}. 
Our apparently different findings
are mutually consistent, in the sense
that there could be covariance in a population, and a hint of anti-covariance on a biased sub-sample of it. Furthermore, since the
\citet{FarahiNat19} sample is formed by more massive clusters than those considered in our work, an unbiased sample of massive clusters would be very useful to establish if the sign of the covariance is mass-dependent.

To summarize, at the typical masses probed by clusters in cluster counts, $13.8<\log M_{500}/M_\odot<14.5$, also probed by the XUCS sample, the found positive covariance between hot and cold baryons offers, at first sight, little help in improving the performances of the two mass proxies taken separately.

\section{Conclusions}
\label{conclusion}

We investigated whether the yet undetermined cause producing clusters of low surface brightness in X-ray also 
induces different galaxy properties, namely richness, richness concentration, width
and location of the red sequence,  colour, luminosity, and dominance of the brightest cluster galaxy.

We used SDSS-DR12 photometry and our analysis factored out the mass dependency to derive trends at fixed cluster mass. 
Clusters of low surface brightness for their mass have cluster richness despite their group-like luminosity.
Gas-poor, low X-ray surface brightness, X-ray faint clusters for their mass, display 25\% lower richnesses for their mass at $4.4\sigma$ level. Therefore, richness and quantities depending on the gas, such as gas fraction, $M_{gas}$, and X-ray surface brightness,  are covariant at fixed halo mass. We do not confirm the hint of anti-covariance between
hot and cold baryons put forward
in previous works for a biased subsample of the cluster population, at least at the intermediate masses explored
in this work.

All the remaining optical properties listed above show no covariance at fixed mass, within the sensitivities allowed
by our data and sample size.  
We conclude that X-ray and optical properties are disjoint, the optical properties not showing 
signatures of those processes involving gas content, apart from the richness-mass scaling relation. 
The covariance between X-ray surface brightness and richness is useful for an effective X-ray follow-up of
low surface brightness clusters because it 
allows us to pre-select them using optical data of survey quality preventing expensive X-ray observations.

In the nearby future, we plan to investigate the subject by means of the comparison with state-of-the-art simulations, by improving the precision of some of our measurements such as concentration with the use of deeper data and by considering a larger number of cluster optical properties. 

\section*{Acknowledgements}
SA with pleasure thanks Bill Forman and Ming Sun for useful discussions on this work. 

\section*{Data availability}

The data underlying this article will be shared on reasonable request to the corresponding author.



\begin{table*}
\caption{XUCS sample. The table lists cluster id (1), cluster coordinates (2-3), redshift (4), followed by values derived in this work: richness measures in $r_{500}$ and in half $r_{500}$ (5-6); concentration measures from the ratio of richness in small and intermediate aperture (7); colour distribution peak and dispersion, as derived in Sect. \ref{sigmacol} (8-9); BCG absolute magnitude and difference between the BCG and the second brightest galaxy (10-11); BCG rest frame colour (12). }             
\label{table:1}      
\centering          
\begin{tabular}{c c c c c c c c c c c c}     
\hline\hline 
\\
 ID & $RA$ & $DEC$ & $z$ & $n_{r500}$ & $n_{r500/2}$  & $\frac{n_{r500/2}}{n_{r500}}$ & $(g-r)$ peak & $\sigma_{col}$ & $M_{BCG}$ & $\Delta M_{BCG}$ & $(g-r)_{rf}$\\
(1) & (2) & (3) & (4) & (5) & (6) & (7) & (8) & (9) & (10) & (11) & (12)\\
\hline                    
\\
CL1001 & 208.256 &  5.134 & 0.079 & 47$\pm$7 & 26$\pm$5 & 0.55 & 0.93 & 0.04$\pm$0.00 & -23.7 & 0.7 & 0.82\\ 
CL1009 & 198.057 & -0.974 & 0.085 & 16$\pm$5 & 12$\pm$4 & 0.75 & 0.97 & 0.03$\pm$0.01 & -23.6 & 1.3 & 0.79\\
CL1011 & 227.107 & -0.266 & 0.091 & 34$\pm$6 & 14$\pm$4 & 0.41 & 0.94 & 0.06$\pm$0.01 & -23.2 & 0.6 & 0.85\\
CL1014 & 175.299 &  5.735 & 0.098 & 31$\pm$6 & 18$\pm$4 & 0.58 & 0.97 & 0.06$\pm$0.02 & -23.4 & 1.1 & 0.86\\
CL1015 & 182.57  &  5.386 & 0.077 & 18$\pm$4 &  9$\pm$3 & 0.50 & 0.93 & 0.02$\pm$0.01 & -23.7 & 1.8 & 0.85\\
CL1018 & 214.398 &  2.053 & 0.054 & 12$\pm$4 &  8$\pm$3 & 0.67 & 0.85 & 0.03$\pm$0.01 & -23.3 & 1.7 & 0.88\\
CL1020 & 176.028 &  5.798 & 0.103 & 29$\pm$6 & 13$\pm$4 & 0.45 & 1.00 & 0.05$\pm$0.01 & -23.8 & 1.5 & 0.79\\
CL1030 & 206.165 &  2.86  & 0.078 & 22$\pm$5 & 14$\pm$4 & 0.64 & 0.91 & 0.08$\pm$0.03 & -22.9 & 0.9 & 0.89\\
CL1033 & 167.747 &  1.128 & 0.097 & 17$\pm$5 &  8$\pm$3 & 0.47 & 0.97 & 0.05$\pm$0.01 & -21.9 & 0.1 & 0.81\\
CL1038 & 179.379 &  5.098 & 0.076 & 25$\pm$6 & 12$\pm$4 & 0.48 & 0.93 & 0.03$\pm$0.00 & -23.3 & 0.7 & 0.81\\
CL1039 & 228.809 &  4.386 & 0.098 & 44$\pm$8 & 25$\pm$5 & 0.57 & 0.98 & 0.03$\pm$0.00 & -23.1 & 0.2 & 0.84\\
CL1041 & 194.673 & -1.761 & 0.084 & 35$\pm$6 & 24$\pm$5 & 0.69 & 0.95 & 0.05$\pm$0.01 & -23.8 & 1.5 & 0.82\\
CL1047 & 229.184 & -0.969 & 0.118 & 30$\pm$6 & 10$\pm$3 & 0.33 & 1.03 & 0.03$\pm$0.01 & -23.2 & 0.9 & 0.82\\
CL1052 & 195.719 & -2.516 & 0.083 & 28$\pm$6 &  9$\pm$3 & 0.32 & 0.95 & 0.04$\pm$0.01 & -21.7 & 0.1 & 0.82\\
CL1067 & 212.022 &  5.418 & 0.088 & 13$\pm$4 &  4$\pm$2 & 0.31 & 0.97 & 0.10$\pm$0.04 & -23.4 & 1.7 & 0.82\\
CL1073 & 170.726 &  1.114 & 0.074 & 27$\pm$6 & 11$\pm$4 & 0.41 & 0.92 & 0.04$\pm$0.01 & -22.8 & 0.5 & 0.78\\
CL1120 & 188.611 &  4.056 & 0.085 & 14$\pm$4 &  8$\pm$3 & 0.57 & 0.93 & 0.08$\pm$0.02 & -22.9 & 0.7 & 0.91\\
CL1132 & 195.143 & -2.134 & 0.085 & 10$\pm$4 &  3$\pm$2 & 0.30 & 0.94 & 0.04$\pm$0.01 & -23.1 & 1.1 & 0.82\\
CL1209 & 149.161 & -0.358 & 0.087 & 12$\pm$4 & 10$\pm$3 & 0.83 & 0.94 & 0.06$\pm$0.02 & -23.3 & 1.6 & 0.83\\
CL2007 & 46.572  & -0.14  & 0.109 & 21$\pm$5 & 13$\pm$4 & 0.62 & 1.02 & 0.03$\pm$0.01 & -22.6 & 0.5 & 0.77\\
CL2010 & 29.071  &  1.051 & 0.08  & 31$\pm$6 & 21$\pm$5 & 0.68 & 0.94 & 0.03$\pm$0.01 & -23.8 & 1.5 & 0.84\\
CL2015 & 13.966  & -9.986 & 0.055 & 24$\pm$5 & 13$\pm$4 & 0.54 & 0.89 & 0.06$\pm$0.01 & -23.8 & 1.7 & 0.82\\
CL2045 & 22.887  &  0.556 & 0.079 & 11$\pm$4 &  4$\pm$2 & 0.36 & 0.96 & 0.06$\pm$0.02 & -23.4 & 1.7 & 0.81\\
CL3000 & 163.402 &  54.87 & 0.072 & 27$\pm$6 & 15$\pm$4 & 0.55 & 0.91 & 0.05$\pm$0.01 & -22.4 & 0.2 & 0.81\\
CL3009 & 136.977 &  52.79 & 0.099 & 21$\pm$5 & 10$\pm$3 & 0.48 & 0.97 & 0.04$\pm$0.01 & -23.4 & 1.2 & 0.83\\
CL3013 & 173.311 &  66.38 & 0.115 & 35$\pm$6 & 14$\pm$4 & 0.40 & 1.05 & 0.04$\pm$0.00 & -24.0 & 1.8 & 0.80\\
CL3020 & 232.311 &  52.86 & 0.073 & 15$\pm$4 &  7$\pm$3 & 0.47 & 0.92 & 0.04$\pm$0.01 & -23.6 & 1.3 & 0.79\\
CL3023 & 122.535 &  35.28 & 0.084 & 14$\pm$4 &  9$\pm$3 & 0.64 & 0.96 & 0.05$\pm$0.02 & -23.3 & 1.0 & 0.82\\
CL3030 & 126.371 &  47.13 & 0.127 & 81$\pm$9 & 39$\pm$6 & 0.48 & 1.07 & 0.05$\pm$0.01 & -24.3 & 1.7 & 0.83\\
CL3046 & 164.599 &  56.79 & 0.135 & 51$\pm$8 & 29$\pm$6 & 0.57 & 1.10 & 0.04$\pm$0.00 & -23.7 & 1.0 & 0.81\\
CL3049 & 203.264 &  60.12 & 0.072 & 12$\pm$4 &  8$\pm$3 & 0.67 & 0.93 & 0.05$\pm$0.02 & -21.6 & 0.1 & 0.77\\
CL3053 & 160.254 &  58.29 & 0.073 &  9$\pm$3 &  5$\pm$2 & 0.56 & 0.94 & 0.04$\pm$0.02 & -22.8 & 0.9 & 0.82\\

\\
\hline                  
\end{tabular}
\end{table*}

%
%



\bibliographystyle{mnras}
\bibliography{dogaspoor.bib} 

\bsp
\label{lastpage}
\end{document}